\documentclass[aps,prb,twocolumn,floats]{revtex4}
\newcommand{\YCBCO}{Y$_{1-y}$Ca$_y$Ba$_2$Cu$_3$O$_{7-\delta}${}}
\newcommand{\YCBCOX}{Y$_{1-y}$Ca$_y$Ba$_2$Cu$_3$O$_{6.92}$}
\newcommand{\kk}{{\bf k}}

\usepackage{graphicx}
\begin{document}
\title{Theory of (001) surface and bulk states in Y$_{1-y}$Ca$_y$Ba$_2$Cu$_3$O$_{7-\delta}$}
\author{K. Pasanai$^{1,2}$} \author{W. A. Atkinson$^{2}$}
\email{billatkinson@trentu.ca} \affiliation{$^1$School of Physics,
  Institute of Science, Suranaree University of Technology, 111
  University Ave., Muang District, Nakhon Ratchasima 30000, Thailand}
\affiliation{$^2$Trent University, 1600 West Bank Dr., Peterborough
  ON, K9J 7B8, Canada} \date{\today}
\begin{abstract}
A self-consistent model is developed for the surface and bulk states
of thin \YCBCO (YCBCO) films.  The dispersions of the chain and plane
layers are modelled by tight-binding bands, and the electronic
structure is then calculated for a finite-thickness film.  The dopant
atoms are treated within a virtual crystal approximation.  Because
YCBCO is a polar material, self-consistent treatment of the long range
Coulomb interaction leads to a transfer of charge between the film
surfaces, and to the formation of surface states.  The tight binding
band parameters are constrained by the requirement that the calculated
band structure of surface states at CuO$_2$-terminated surfaces be in
agreement with photoemission experiments.  The spectral function and
density of states are calculated and compared with experiments.
Unlike the case of $\mathrm{Bi_2Sr_2CaCu_2O_8}$, where the surfaces
are believed to be representative of the bulk, the densities of states
at the YCBCO surfaces are shown to be qualitatively different from the
bulk, and are sensitive to doping.  The calculated spectral function
agrees closely with both bulk-sensitive and surface-sensitive
photoemission results, while the calculated density of states for
optimally-doped YCBCO agrees closely with tunneling experiments.  We
find that some density of states features previously ascribed to
competing order can be understood as band structure effects.
\end{abstract}
\maketitle
\section{Introduction}
The density of states (DOS) measured by tunneling experiments in the
$\mathrm{YBa_2Cu_3O_{7-\delta}}$ (YBCO)  family of high
temperature superconductors is complicated.  Several experiments on
superconducting samples have measured densities of states with multiple energy
scales.\cite{Valles1991,Maggio-Aprile1995,Yeh2001,Ngai2007,Das2008,Beyer2009,Fischer2007}
Some experiments find a subgap feature\cite{Yeh2001,Ngai2007} while others do
not,\cite{Maggio-Aprile1995,CrenEPL2000,Das2008,Beyer2009} and all experiments
near optimal doping find satellite features at energies larger than
the gap energy.  Many of these studies find that the spectra change
qualitatively with doping and can vary significantly at different
points on the sample surface.\cite{Ngai2007,Beyer2009} Spectral
features have been interpreted in terms of band
structure,\cite{Ngai2007} competing order,\cite{Beyer2008,Beyer2009}
and coupling to bosonic modes.\cite{Das2008} 

In this work, we explore reasons why the
YBCO  single-particle spectrum is so
complicated, particularly when compared to that of the related
superconductor $\mathrm{Bi_2Sr_2CaCu_2O_8}$
(BSCCO).\cite{Cren2000,Pan2001,Howald2001} In BSCCO, there is a clear
$d$-wave-like gap in the density of states, and it has been possible
to reproducibly extract detailed information about the band
structure\cite{Norman1995,Markiewicz2005,Fischer2007} and
superconducting state.\cite{Yusof1998,Hoogenboom2003} At present,
there is no consensus on how to interpret the tunneling DOS in YBCO.  In this work, we focus on two specific structural
differences between BSCCO and YBCO, namely that YBCO is a polar
material while BSCCO is not, and that YBCO has metallic
one-dimensional CuO chains while BSCCO does not.  We show that the
confluence of these two factors explains some features of the
experimentally-measured DOS.  In particular, our results
suggest that some features that were previously thought to indicate
charge-ordering actually come from peculiarities of the YBCO band
structure.

The polarity of the YBCO unit cell is important for several reasons.
Unlike BSCCO, YBCO has no natural cleavage plane (i.e.\ no plane along
which ionic forces vanish), making it difficult to prepare surfaces
that are clean enough for experiments.  More importantly for this
work, there is a charge transfer between YBCO surfaces as a result of
the electric fields generated by the polar unit cell.  This charge
transfer leads to the formation of surface states that can differ
significantly from states in the bulk.  In contrast, it is widely
believed that the surface layers in BSCCO are representative of the
bulk.

Surface charging in polar crystals has recently become prominent in
the context of $\mathrm{LaAlO_3}$/$\mathrm{SrTiO_3}$
interfaces.\cite{Nakagawa2006,Thiel2006} The essential idea is that,
since the opposite faces of a polar unit cell have opposite charge,
there is a potential difference between them.  In a thin film, the
potential difference between the top and bottom surfaces of the film
is equal to the potential difference across a single unit cell
times the number of unit cells spanning the film.  The potential
difference between the surfaces of the film is thus proportional to
the film thickness, much like a parallel plate capacitor, and is
typically large when the sample is more than a few unit cells thick.
A ``polar catastrophe'' (i.e.\ a divergent electrostatic energy as the
sample becomes macroscopically thick) is avoided by a transfer of
charge between the two surfaces.  This screening charge eliminates the
potential difference across the film, but changes the doping at the
surfaces and leads to the formation of surface states.  The existence
of surface states in YBCO has recently been confirmed by angle resolved
photoemission (ARPES) experiments.\cite{Nakayama2007,Zabolotnyy2007}

The second aspect of YBCO that makes it distinct from BSCCO is the
presence of layers of one-dimensional CuO chains, in addition to the
CuO$_2$ plane bilayers.  The na\"ive view is that these chains carry
charge in parallel to the CuO$_2$ bilayers, but have little direct
impact on the CuO$_2$ layers.  The chains are, therefore, generally
ignored in models of YBCO.  For in-plane transport experiments, this
point of view appears justified since it is possible to eliminate the
effects of the chains (which run parallel to the $b$ axis) by
measuring transport in the $a$-axis direction; however, the chains are
not easily disentangled from most other types of experiment.  For
example, $c$-axis currents (perpendicular to the planes) must pass
through the CuO$_2$ and CuO layers in series, so that the $c$-axis
resistivity is dominated by the Fermi surface mismatch between plane
and chain layers.\cite{Atkinson1997a} As another example, CuO chains
have been argued to cause an anomalous vortex core expansion at low
magnetic fields in YBCO,\cite{Atkinson2008} which is connected to a
small superconducting energy scale in the chains.\cite{Whelan2000}  In a similar vein,
we find in this work that the effects of the chains are subtle, but
are important for understanding some details of the density of states
in the CuO$_2$ layers.

In this work, we calculate the tunneling DOS for a
tight-binding model of YBCO that is based on recent experimental ARPES
measurements of the band structure.  There have, over the years, been
many attempts to measure the YBCO spectrum using ARPES but, for
reasons discussed above, it is difficult to do reliably.  Some of the
first successful measurements were made by Schabel et
al,\cite{Schabel1998a,Schabel1998b} who found a complicated set of
bands, not all of which could be easily related to bands predicted by
density functional theory (DFT) calculations.  Later work by Lu et
al\cite{Lu2001} identified an anisotropic spectrum consistent with the
presence of CuO chains, and measurements in $\mathrm{YBa_2Cu_4O_8}$
confirmed the existence of a pair of chain Fermi surfaces in that
material.\cite{Kondo2007} Recent experiments by Zabolotnyy et
al,\cite{Zabolotnyy2007} and Nakayama et al\cite{Nakayama2007} 
mapped out the Fermi surfaces of the surface states in some detail.
%Chain Fermi surfaces were imaged along with CuO$_2$ plane
%Fermi surfaces in \onlinecite{Zabolotnyy2007} and \onlinecite{Nakayama2007}, 
Most recently Okawa et al\cite{Okawa2009} have succeeded in imaging
states in the bulk, and were able to partially map the Fermi surface and
superconducting gap near the middle of the Brillouin zone.

The goal of this work is to develop a self-consistent model for surface and
bulk states in \YCBCO, and to use
this model to understand details of the tunneling DOS.
We remark that YBCO is sufficiently three-dimensional that it is not
possible to model surface states by considering a single isolated
CuO$_2$ layer, as is frequently done to model BSCCO.  Instead, we
develop a model for a $c$-axis oriented film of thickness $N_c$ unit
cells.

The approach we take is phenomenological.  In Sec.~\ref{sec:method},
we develop a model Hamiltonian for YBCO from least-squares fits of a
tight-binding dispersion to the surface states measured in Ref.~\onlinecite{Nakayama2007}.  We focus on optimal and overdoped YBCO,
where we hope to avoid complications arising from competing phases or
quantum critical points.  The ARPES spectrum gives us the
two-dimensional (2D) dispersion curves for the surface states; we assume
that these dispersions are rigid, meaning that dispersions for bulk
CuO$_2$ layers may be obtained by shifting the surface
dispersions up or down in energy.  
This process allows us to
infer the structure of the bulk bands based on the measured surface
states.  We find that the inferred bands agree with 
the measured bulk Fermi surface from Ref.~\onlinecite{Okawa2009}.
 We then perform self-consistent calculations
for the electrostatic potential in a finite-thickness film;  this gives us the band-bending
profile near the surfaces.  
  We add superconductivity to the CuO$_2$ layers by hand,
using a phenomenological relationship between the magnitude of the
order parameter and the charge density.  In this way, it is found that
the order parameter is smaller on CuO$_2$-terminated surfaces than in
the bulk.  Finally, we calculate the superconducting DOS
and normal state spectral functions for both surface and bulk layers.
We discuss the results of these calculations in Sec.~\ref{sec:result}.
In this section, we conclude that, unlike in BSCCO, the density of
states is sensitive to details of the band structure, and that these
details naturally explain some of the features seen experimentally.
%-----------------------end introduction--------------------------------------------------------------

\begin{figure}
\includegraphics[width=0.9\columnwidth]{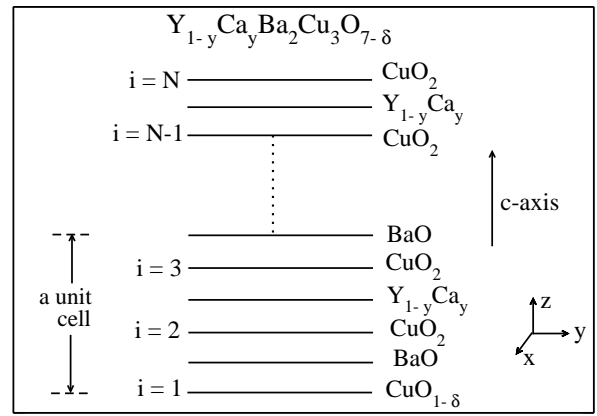}
\caption{Model of a Ca-doped YBCO thin film.  The film is
  $N_c$ unit cells thick, and each unit cell contains two
  superconducting CuO$_2$ and one metallic CuO layer, for a total of
  $N=3N_c$ conducting layers.  The conducting layers are
  labelled $i=1,\ldots,N$, with $i=1$ corresponding to the
  chain-terminated surface, and $i=N$ corresponding to the CuO$_2$
  plane-terminated surface.  The CuO$_2$ and CuO layers are described
  by two- and one-dimensional dispersions $\xi_{ip}(\kk)$ and
  $\xi_{ic}(\kk)$ respectively, where $\kk=(k_x,k_y)$ and are coupled
  by interlayer hopping matrix elements $t_{\perp p}$ (plane-plane) and
  $t_{\perp c}$ (chain-plane).  The nonconducting layers are not
  included in the band structure calculations, although the Y/Ca
  layers are included in calculations of the electrostatic
  potential. The BaO layers are electrically neutral and are ignored.
  Results in this work are shown for $N_c=10$ unit cells.}
\label{fig:model}
\end{figure}

%------------------------ Method ------------------------------------------------------------------------
\section{Method}
\label{sec:method}
%\subsection{Tight-binding Hamiltonian}
%\label{sub:Tight_binding_Hamiltonian}

We consider a thin film consisting of N$_c$ unit cells, stacked along
their $c$-axis, as illustrated in Fig.~\ref{fig:model}. Each unit cell
contains three conducting layers (two CuO$_2$ planes and one CuO chain
layer) and three insulating layers (a Y$_{1-y}$Ca$_y$ layer and two BaO layers).
The total number of conducting layers is therefore $N=3N_c$.
Experimentally, it is known that YBCO cleaves at
the BaO layer, so that the top conducting layer may be either a CuO
chain layer or a CuO$_2$ plane layer.  To study both cases, we assume
that the first layer ($i=1$) is composed of CuO chains, and the last
layer ($i=N$) is a CuO$_2$ plane.

Of the three nonconducting layers, two (the BaO layers) are nominally
neutral and are not explicitly considered in the model.  The yttrium and calcium
atoms are nominally in the Y$^{3+}$ and Ca$^{2+}$ states and are retained in
calculations of the long-ranged Coulomb potential.

Of the three conducting layers, it is assumed that only the CuO$_2$
planes are intrinsically superconducting, and that the chains are
superconducting because of their proximity to the planes.  The
proximity model for chain superconductivity has been discussed at
length elsewhere.\cite{Tachiki1990,ODonovan1997,Atkinson1999,Morr2001} The essential idea is that
single-electron hopping between the chain and plane layers leads to
mixing of the electronic wavefunctions, and to induced pairing in the
chain layer.

The mean-field Hamiltonian of the system is
\begin{equation}
H = \sum_\kk \hat C_\kk^\dagger H_\kk \hat C_\kk
\end{equation}
where $\hat C_\kk^\dagger = [\hat c_{1\kk\uparrow}^\dagger, \hat
  c_{1-\kk\downarrow},\ldots,\hat c_{N\kk\uparrow}^\dagger, \hat
  c_{N-\kk\downarrow}]$ and $\hat c_{i\kk\sigma}^\dagger$ is the
creation operator for a spin-$\sigma$ electron in the conducting layer
$i$ with 2D wave vector $\kk = (k_x,k_y)$. $H_\kk$ is
given by
\begin{widetext}
\begin{equation}
\label{eq:hamiltonian}
H_\kk = \left[ {\begin{array}{*{15}c}
   {\xi _{1c} (\kk)} & 0 & {t_{ \bot c} } &  &  &  &  &  &   \\
   0 & { - \xi _{1c} (\kk)} & 0 & { - t_{ \bot c} } &  &  &  &  &   \\
   {t_{ \bot c} } & 0 & {\xi _{2p} (\kk)} & {\Delta _{2\kk} } & {t_{ \bot p} } &  &  &  &   \\
    & { - t_{ \bot c} } & {\Delta _{2\kk} } & { - \xi _{2p} (\kk)} & 0 & { - t_{ \bot p} } &  & 0 &   \\
    &  & {t_{ \bot p} } & 0 & {\xi _{3p} (\kk)} & {\Delta _{3\kk} } &  &  &  \\
    &  &  & { - t_{ \bot p} } & {\Delta _{3\kk} } & { - \xi _{3p} (\kk)} &  &  &   \\
    & 0 &  &  &  &  &  \ddots &  & { - t_{ \bot p} }  \\
    &  &  &  &  &  &  & \xi _{Np} (\kk) & {\Delta _{N\kk} }  \\
    &  &  &  &  &  & { - t_{ \bot p} } & {\Delta _{N\kk} } & { - \xi _{Np} (\kk)}  \\

 \end{array} } \right]
\end{equation}
\end{widetext}
where $\xi_{ic(p)}(\kk)$ is the 2D chain (plane) energy
dispersion in conducting layer $i$. The matrix elements $t_{\perp c}$
and $t_{\perp p}$ are for plane-chain and plane-plane hopping
respectively.  We remark that, in BSCCO, the plane-plane hopping
matrix element has a strong $\kk$-dependence; ARPES
experiments\cite{Okawa2009} and DFT calculations\cite{Andersen1995}
agree that the $\kk$-dependence in YBCO is weaker, and we have
neglected it here as a way of reducing the number of fitting
parameters in our model.  

For the  intralayer dispersions, we use
tight-binding models with nearest neighbor and third-nearest neighbor
hopping for the chains and planes respectivley:
\begin{eqnarray}
\xi_{ic}(\kk)  &=& -2t_c\cos k_y + \Phi_{i} 
\label{eq:2Dchain} \\
\xi_{ip}(\kk) &=& -2t_p[\cos k_x + \cos k_y + 2t^\prime\cos k_x\cos k_y \nonumber \\
             &+& t^{\prime\prime}(\cos 2k_x + \cos 2k_y)  \nonumber \\
             &+ & 2t^{\prime\prime\prime}(\cos 2k_x\cos k_y + \cos 2k_y\cos k_x )]  \nonumber \\
			 &+& \Phi_{i}
\label{eq:2Dplane}
\end{eqnarray}
where the values of
$t_c,~t_p,~t^\prime,~t^{\prime\prime},~t^{\prime\prime\prime}$, and
$t_{\perp p}$  are shown in table \ref{table:param}.
The potential $\Phi_i$ includes short and long range Coulomb
interactions, as well as offset potentials for the plane and chain
layers, and the chemical potential.  The method of calculating
$\Phi_i$ is discussed below.

It is known that the superconducting order parameter in the CuO$_2$
planes has $d_{x^2-y^2} + s$ symmetry\cite{Ye2007} as a result of the
orthorhombic crystal structure. 
From experiments,\cite{Smilde2005} the s-wave component
is approximately $15\%$ of the d-wave component, and we assume that this
ratio holds at both the surfaces and in the bulk.  Denoting the d-wave
and s-wave components in layer $i$ by
 $\Delta_{id}$ and $\Delta_{is}$ respectively, we have
\begin{equation}
\label{eq:oderparameter}
\Delta_{i\kk} = \left \{ 
\begin{array}{ll}
  \frac{\Delta_{id}}{2}[\cos (k_x)-\cos (k_y)] +
  \Delta_{is}, & i\in\mbox{plane} \\ 
0, & i\in\mbox{chain}
\end{array}
\right.
\end{equation}
Note that, although the order parameter is zero in the chains, 
superconductivity is induced by the proximity effect. We adopt the 
phenomenological expression for $\Delta_{id}$ in the CuO$_2$ layers
\begin{equation}
\Delta_{id} = \frac{100[n_{i} - n_\mathrm{min}]}{1-n_\mathrm{min}} \mbox{ meV},
\label{eq:simple_gap}
\end{equation}
where $n_{i}$ is the charge density in plane $i$ and
$n_\mathrm{min}=0.7$.  We expect (\ref{eq:simple_gap}) to be valid on
the overdoped side of the phase diagram.  At optimal doping (with
planar charge density $n_p\approx 0.84$ in the bulk), this gives a
$d$-wave order parameter of magnitude $46$ meV, which is close to
values inferred from recent ARPES measurements.\cite{Okawa2009}
Equation (\ref{eq:simple_gap}) also implies that the gap vanishes when
the electron concentration is less than $n_\mathrm{min}$.  The value
of $n_\mathrm{min}$ is not well-known and is probably
material-dependent (depending, for example, on the level of
doping-related disorder); the canonical form for
$\mathrm{La_{2-x}Sr_xCuO_4}$ gives $n_\mathrm{min}=0.73$, but
tunneling experiments\cite{Ngai2007} on \YCBCO{} suggest that
superconductivity is still present at this doping level in YBCO.  Our
choice seems reasonably consistent with these experiments.

The potentials in (\ref{eq:2Dchain}) and (\ref{eq:2Dplane})
are calculated from a  self-consistent mean-field treatment.
We write
\begin{equation}
\Phi_{i}= \left \{ \begin{array}{ll}  
\phi_{i}  + \epsilon_{p},& i\in\mbox{plane} \\
\phi_{i}  + \epsilon_{c},& i\in\mbox{chain}
\end{array} \right .
\end{equation}
where $\phi_i$ is the electrostatic potential, and $\epsilon_{c(p)}$
includes the chemical potential and the energy of the chain (plane)
tight binding orbitals.  We determine $\epsilon_p$ and $\epsilon_c$ by
specifying the bulk plane and chain charge densities at optimal doping
($\delta=0.08$), which we take to be $n_p=0.84$ and $n_c=0.48$
electrons per 2D unit cell respectively.  Note that, once $n_p$ is
chosen, $n_c$ is set by the constraint of charge neutrality, given by
Eq.~(\ref{eq:dop}) below.  Our self-consistent calculations then find
$\epsilon_c-\epsilon_p=3.176$ eV at optimal doping. We assume that the
tight binding orbitals are not modified by doping, so that
$\epsilon_c-\epsilon_p$ is held constant througout this work.  In our
calculations chemical doping modifies the band structure only through
the electrostatic potential $\phi_i$.

The Coulomb potential $\phi_i$ is then calculated self-consistently
within the Hartree approximation, under the assumption that the charge
is uniformly distributed within each layer.  For the planar geometry
shown in Fig.~\ref{fig:model},
\begin{equation}
  \phi_i = \left \{
\begin{array}{ll} 
-\kappa \sum_j \sigma_j|z_i-z_j| + U\frac{n_i}{2}, & i\in\mbox{plane} \\
-\kappa \sum_j \sigma_j  |z_i-z_j|, & i\in\mbox{chain} 
\end{array}
\right .
\label{eq:potential}
\end{equation}
where the Y$_{1-y}$Ca$_y$ layers are implicitly included in the sum over $j$ and 
\begin{equation}
\kappa = \frac{2\pi e^2d_z}{\epsilon a_0^2}.
\end{equation}
In (\ref{eq:potential}), $U$ is the intraorbital Coulomb potential for
the CuO$_2$ layers, which we take to be 4~eV.  
The total 2D charge density in layer $i$ is
$\sigma_i=Z_i-n_i$, where $Z_i$ is the charge density of the ionic
cores.  Here, $n_i$ is measured relative to the Cu$^{3+}$ and O$^{2-}$
states, so that 
\begin{equation}
\sigma_j =\left \{ \begin{array}{ll} 1+2\delta-n_{j}, &j\in\mbox{CuO$_{1-\delta}$ chain} \\
-1-n_{i}, & j\in\mbox{CuO$_2$ plane} \\
3-y, & j\in\mathrm{Y_{1-y}^{3+}Ca_y^{2+}}\mbox{ layer}
\end{array} \right .
\end{equation}
  All charge densities are in units of $e/a_0^2$, where
$a_0\approx 4$ \AA{} is the 2D lattice constant.  Electrical
neutrality requires that
\begin{equation}
n_c + 2n_p = 2 +2\delta-y.
\label{eq:dop}
\end{equation}
In Eq.~(\ref{eq:potential}), $z_i$ is the $z$-coordinate of layer $i$,
 in units of the $c$-axis lattice constant
$d_z\approx 12$~\AA.  Within a unit cell, the layers are at $z=0$
(chain), $z=0.354 $ (first plane), $z=0.5$ (Y layer), and $z =
0.646$ (second plane).\cite{Bottger1996} The weak doping dependence of
these values is ignored here.  The dielectric constant
$\epsilon$ in (\ref{eq:potential}) is not well known, but is believed
to be around $\epsilon=20$, which is the value taken here.

For a given potential $\Phi_i$, the charge density in layer $i$ is
found from the eigenvalues $E_{\alpha,\kk}$ and eigenstates
$\Psi_{\alpha,\kk}(i,\sigma)$ of $H_\kk$ via
\begin{equation}
  n_i = \frac{2}{N_k}\sum_\kk\sum_{\alpha=1}^{2N}
  |\Psi_{\alpha,\kk}(i,\uparrow)|^2 f(E_{\alpha,\kk})
\end{equation}
where $f(x)$ is the Fermi function, the factor of 2 is for spin, and $N_k$ is the number of k-points in the sum.  The band index $\alpha$
ranges from 1 to $2N$ because the number of bands in the film is equal to twice (including spin) the number of conducting layers.  The
updated charge density is used to re-calculate $\Phi_i$, which is then
used in the next iteration for $n_i$.  The iterations proceed until
the difference between $n_i$ in two consecutive iterations is less
than $10^{-5}$.  To reduce the computational workload, $\Phi_i$ is
calculated in the normal state.

\begin{table}
\caption{Parameters for the tight binding model for YBCO.}
\begin{tabular}{|c|c|}
\hline
Parameter & Value  \\
\hline
$t_p$  & 105 meV\\
$t^\prime$     & -0.277\\
$t^{\prime\prime}$    & 0.234\\
$t^{\prime\prime\prime}$   & -0.042\\
$t_c$  & 500 meV\\
$t_{\perp p}$ & 61 meV\\
$t_{\perp c}$ & 1.1$t_{\perp p}$ \\
\hline
\end{tabular}
\label{table:param}
\end{table}

We finish this section with a brief discussion of the fitting
procedure used to get the model parameters shown in
Table~\ref{table:param}.  As discussed above, we fit the 2D
dispersions for the CuO$_2$ layers to the surface states measured in
Ref.~\onlinecite{Nakayama2007}.  In order to avoid complications from
the chains, we fit the energy spectrum for an isolated CuO$_2$ bilayer
to the measured bands in regions of the Brillouin zone far from the
chain Fermi surface.  The model bilayer has bonding and antibonding
bands, with energies $\xi^\pm(\kk) = \xi_{p}(\kk) \pm t_{\perp p}$,
which allows us to determine
$t_p,~t^\prime,~t^{\prime\prime},~t^{\prime\prime\prime}$, and
$t_{\perp p}$.  One potential difficulty with this fitting process is
that it assumes that the electrostatic potential is the same in the top
two CuO$_2$ layers, meaning that we attribute the experimental bilayer
splitting entirely to $t_{\perp p}$.  If we allow for a potential
difference $\Delta \phi$ between the CuO$_2$ layers making up the
bilayer (due to band bending at the surfaces), then $\xi^\pm(\kk) = \xi_{p}(\kk) \pm \sqrt{(\Delta\phi/2)^2
  + t_{\perp p}^2}$.  Self-consistent calculations reported in the
next section suggest $\Delta \phi \sim 10$ meV while the measured
$t_{\perp p}\approx 60$ meV; it follows that $\Delta\phi$ modifies the
band energies by less than 1\% and can safely be neglected.

There are, at present, no reliable measurements of the chain band
structure.  We therefore assume that strong correlations are not
significant in the CuO chains (which are roughly quarter-filled), and
that the chain bandwidth of $\sim 2$ eV from DFT is
reasonable.\cite{Andersen1995} This gives the intrachain hopping
matrix element $t_c = 500$ meV.  The hardest parameter to establish is
$t_{\perp c}$, the chain-plane coupling. We show results for multiple
values of $t_{\perp  c}$ in Sec.\ref{sec:LDOS_ARPES}, and find that
the experimental density of states is reasonably well fit for
$t_{\perp c}=1.1t_{\perp p}$ for YBa$_2$Cu$_3$O$_{6.92}$.

\begin{figure}
\includegraphics[width=\columnwidth]{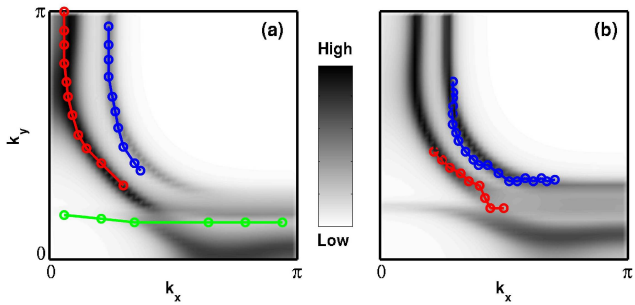}
\caption{(Color online)  Comparison of the self-consistently calculated spectral function at
$\varepsilon_F$ to the experimentally measured Fermi surface (circles).  Results are
for (a) surface states (data from Ref.~\protect\onlinecite{Nakayama2007}) and (b)
bulk states (data from Ref.~\protect\onlinecite{Okawa2009}).}
\label{fig:fitting}
\end{figure}
We show the results of our fitting procedure in Fig.~\ref{fig:fitting}.  In this figure, we compare calculated surface and bulk spectral functions at the Fermi energy with the surface and bulk spectral functions measured by ARPES.  The calculations are in good agreement with the experiments, which suggests that the assumptions made in developing our model are reasonable. 

%----------------- result ------------------------------------------------------------------------------
\section{Results and Discussion}
\label{sec:result}
\subsection{Self-consistent potential, charge density, and superconducting
order parameter}
\label{sec:potential}

\begin{figure}
\includegraphics[width=\columnwidth]{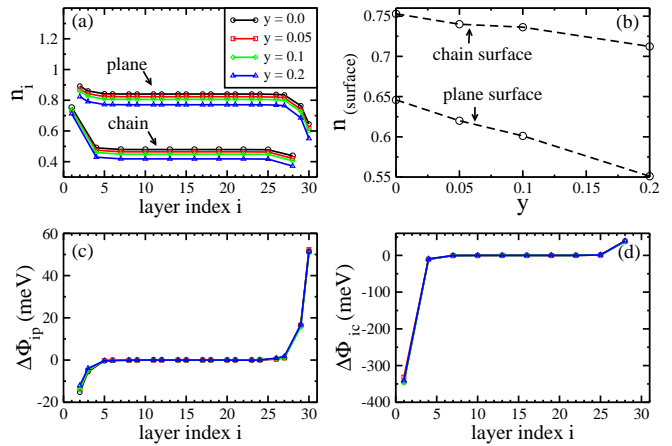}
\caption{(Color online) Self-consistent solutions for the charge
  density and electrostatic potential in \YCBCOX.  Charge density is
  shown (a) for all CuO chain and CuO$_2$ plane layers in the thin
  film as a function of the layer index $i$, and (b) at the chain
  ($i=1$) and plane ($i=30$) surfaces as a function of $y$.  The
  potential difference $\Delta \Phi_{ip(c)} = \Phi_i - \Phi_{p(c)}$,
  between the potential in layer $i$ and the potential in the bulk
  planes (chains), is shown in (c) and (d).}
\label{fig:charge_on_layer}
\end{figure}

Figure \ref{fig:charge_on_layer} shows the results of self-consistent
calculations for the charge density and electrostatic potential.  We
take the specific case of \YCBCOX, which corresponds to optimal doping
when $y=0$ and to overdoping when $y>0$.  The self-consistently
determined charge density is shown in
Fig.~\ref{fig:charge_on_layer}(a) as a function of layer index $i$.
(Recall that the layer indices $i=1$ and $i=30$ label the CuO
chain-terminated and the CuO$_2$ plane-terminated surfaces
respectively.)  As discussed in the introduction,
Fig.~\ref{fig:charge_on_layer}(a) shows that there is charge
transfer from the plane-terminated surface to the chain-terminated
surface.  This charge screens the electric field produced by the polar
unit cells, so that the electric potential is constant across the thin
film except near the surfaces.  This is shown in
Fig.~\ref{fig:charge_on_layer}(c) and (d). In these figures, we have
plotted the difference
\begin{equation}
\Delta \Phi_{ip(c)} = \Phi_i - \Phi_{p(c)},
\end{equation}
between the potential in layer $i$ and the potential for a plane
(chain) in the bulk in order to make the comparison between different
$y$ values simpler.  $\Delta \Phi_{ip(c)}$ is nonzero within a
screening length of the surfaces and is positive (negative) at the
plane (chain) surface. The charge density on the plane (chain) surface
is correspondingly smaller (larger) than in the bulk as shown in
Fig.~\ref{fig:charge_on_layer}(a).

We note that $\Delta \Phi_{1c}$, the potential shift at the
chain-terminated surface, is roughly seven times larger than $\Delta
\Phi_{30p}$, the potential shift at the plane-terminated surface.
This ratio is approximately the same as the ratio of the plane and
chain densities of states (in the normal state); because of the small
density of states in the chains, a large chemical potential shift is
required at the chain-terminated surface to accommodate the charge
transferred from the plane-terminated surface.

Figure~\ref{fig:charge_on_layer} also shows the effect of Ca
substitution.  Despite the proximity of the Ca ions to the CuO$_2$
layers, $n_i$ changes by roughly the same amount in both the CuO$_2$
and CuO layers.  The one notable exception to this is at the chain
surface, where the electron concentration changes by roughly half as
much as in the bulk [Fig.~\ref{fig:charge_on_layer}(b)].

\begin{figure}
\includegraphics[width=\columnwidth]{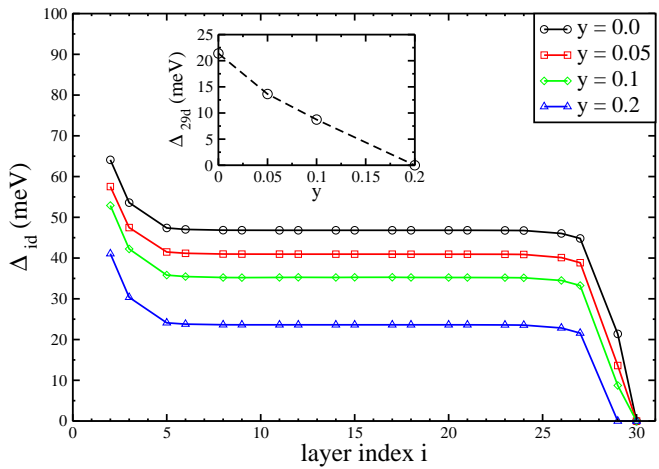}
\caption{(Color online) Superconducting order parameter as a function of
layer index for \YCBCOX.  The figure shows
the $d$-wave component of the order parameter, which is nonzero in the
CuO$_2$ layers only.  The magnitude of $\Delta_{id}$ is given by Eq.~(\protect\ref{eq:simple_gap}).   {\em Inset:} Superconducting order parameter near the plane 
surface ($i=29$) as a function of Ca doping. Note that $\Delta_{30d}=0$.
 }
\label{fig:gap_vs_nl}
\end{figure}

Figure \ref{fig:gap_vs_nl} shows the $d$-wave superconducting order
parameter $\Delta_{id}$ as a function of layer index $i$ in the
CuO$_2$ plane layers, and for different levels of Ca doping.  Because
$\Delta_{id}$ is calculated phenomenologically from
Eq.~(\ref{eq:simple_gap}), it follows $n_i$.  Thus, $\Delta_{id}$ is
larger than in the bulk near the chain surface and smaller than in the
bulk near the plane surface.  In our calculations, $\Delta_{id}$
actually vanishes at the CuO$_2$ surface layer ($i=30$), although
there is a spectral gap due to proximity coupling to the subsurface
layers.  Note that recent ARPES
experiments,\cite{Zabolotnyy2007,Nakayama2007} found that the surface
is nonsuperconducting, but tunneling
experiments\cite{Valles1991,Maggio-Aprile1995,Yeh2001,Ngai2007,Das2008,Beyer2009,Fischer2007}
found a clear superconducting gap at the surface.  We will show in the
next section that the size of the superconducting gap at the CuO$_2$
surface is sensitive to how the surface is prepared.

%----------------- LDOS and ARPES--------------------------------------------------------
\subsection{Density of states and spectral function}
\label{sec:LDOS_ARPES}

The main results reported in this work are for the layer-dependent density of states and spectral 
function. The density of states at energy 
$\omega$ in layer $i$ is given by
\begin{eqnarray}
\rho_i(\omega) &=& \frac{1}{N_\kk} \sum_{\kk} A_i(\kk,\omega)
% \sum_{\alpha=1}^{2N} 
%[|\Psi_{\alpha,\kk}(i,\uparrow)|^2\delta(\omega-E_{\alpha,\kk}) \nonumber \\
%&+& |\Psi_{\alpha,\kk}(i,\downarrow)|^2\delta(\omega+E_{\alpha,\kk})]
\end{eqnarray}
where $N_\kk$ is the total number of $\kk$-points,
and $A_i(\kk,\omega)$ is spectral function  in layer $i$, 
\begin{eqnarray}
A_i(\kk,\omega) &=&
\sum_{\alpha=1}^{2N} [ |\Psi_{\alpha,\kk}(i,\uparrow)|^2 \delta (\omega - E_{\alpha,\kk})
\nonumber \\
&&+ |\Psi_{\alpha,\kk}(i,\downarrow)|^2 \delta (\omega + E_{\alpha,\kk}) ].
\end{eqnarray}
  Note that the eigenstates 
$\Psi_{\alpha,\kk}(i,\sigma)$ and eigenenergies $E_{\alpha,\kk}$
describe electrons ($\sigma=\uparrow$)
or holes ($\sigma=\downarrow$) depending on the spin index.
In this work, all results for $A_i(\kk,\omega)$ are shown for the normal state at
$\omega=\varepsilon_F$,
while results for $\rho_i(\omega)$ are shown in the superconducting state.

%------------------------------surface and bulk LDOS------------------------------------- 
\begin{figure}
\includegraphics[width=\columnwidth]{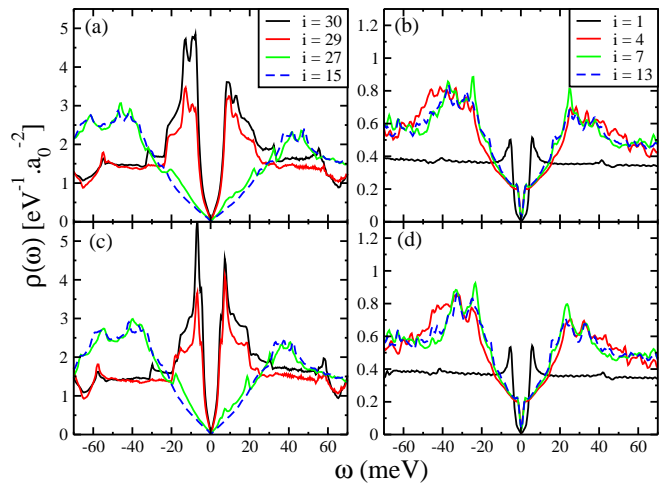}
\caption{(Color online) Surface and bulk density of states for
  \YCBCOX.  Results are shown for y = 0.0 [(a) and (b)] and for y =
  0.05 [(c) and (d)].  DOS is shown for CuO$_2$ planes [(a) and (c)] and
  [(b) and (d)] CuO chains. Model parameters are given in
  Table~\ref{table:param}. }
\label{fig:simpledos}
\end{figure}

Results for \YCBCOX{} with $y=0$ and $y=0.05$ are shown in
Fig.~\ref{fig:simpledos}.  This figure shows that the density of
states at the surfaces is very different from in the bulk, for both
the CuO$_2$ planes and CuO chains.  Notably, the $d$-wave gap is
significantly smaller at the surfaces than in the bulk.  In the
CuO$_2$ layers, this reflects the suppression of $\Delta_{id}$ near
the surface due to the decreased electron density in the surface
layers.  This is consistent with bulk-sensitive ARPES experiments that
found a gap of order 42 meV,\cite{Okawa2009} and surface-sensitive
tunneling experiments, which consistently find that  the gap is
$\lesssim 25$
meV.\cite{Valles1991,Maggio-Aprile1995,Yeh2001,Ngai2007,Das2008,Beyer2009,Fischer2007}
In the chain layers, the reduced gap at the surface reflects the
reduced proximity coupling at the chain surface, in part because the
chain has only one nearest neighbor CuO$_2$ plane.  Our calculation
shows that the DOS obtains its bulk value within a few layers of
either surface.

\begin{figure}
\includegraphics[width=\columnwidth]{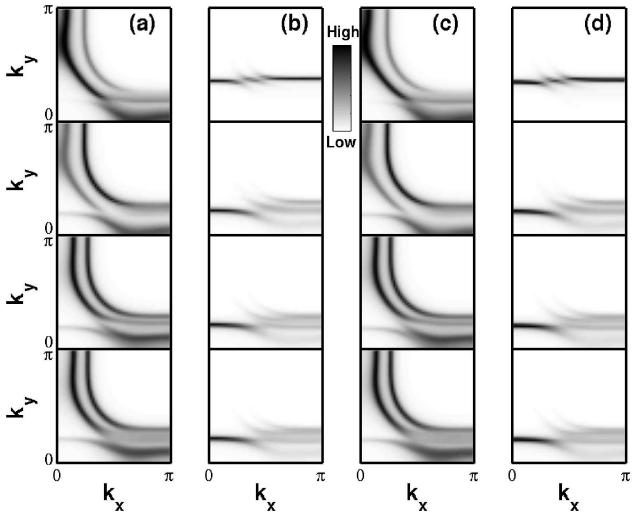}
\caption{Spectral function at the surfaces and in the bulk of \YCBCOX.
  Columns show $A(\kk,\varepsilon_F)$ for $y=0$ for (a) planes and (b)
  chains, and for $y=0.05$ for (c) planes and (d) chains.  Rows (top
  to bottom) show layers $i=30$, 29, 27, 15 (planes) and $i=1$, 4, 7,
  13 (chains); thus, surface states are shown in the top row, states
  deep in the bulk are shown in the bottom row.  }
\label{fig:simpledos_1}
\end{figure}

We note that, relative to the conventional model of a single-layer
$d$-wave superconductor, the DOS in Fig.~\ref{fig:simpledos} shows a
lot of structure.  In particular, the CuO$_2$ surface (for $y=0$) has
a main gap of about 10 meV, whose coherence peaks are split into pairs
of closely-spaced peaks, and a satellite ``shoulder" at about 20 meV.
Shoulder features have been commonly observed in tunneling
experiments, and have been attributed to pairing,\cite{Ngai2007} and
to competing order.\cite{Beyer2009} In our calculations, this
structure comes from the interplay of pairing and band structure
effects, namely the mixing of chain and plane states resulting in an
orthorhombic distortion of the Fermi surfaces.  This is illustrated in
Fig.~\ref{fig:simpledos_1}, which shows the layer-resolved spectral
function $A_i(\kk,\varepsilon_F)$ at the Fermi energy.  We see from
this figure that chain-plane coupling strongly distorts the Fermi
surfaces in the $(\pi,0)$ region of the Brillouin zone.  This
distortion is particularly important for the CuO$_2$ layers, because
there is a van Hove singularity near the $(\pi,0)$ point in the
undistorted spectrum.  The DOS is therefore sensitive to small changes
in the Fermi surface shape, caused either by chain-plane coupling, by
doping, or by band-bending (changes in the electrostatic potential)
near the surfaces.  Thus, the addition of 5\% Ca changes the electron
density by only $\sim 0.02$ electrons per CuO$_2$ plaquette but
qualitatively changes the shape of the coherence peaks [compare
Fig.~\ref{fig:simpledos}(a) and (c)]; in Fig.~\ref{fig:simpledos_1}, we
see that the main effect on the CuO$_2$ surface layer ($i=30$) is
indeed near the $(\pi,0)$ point, where the spectral weight is reduced
by Ca doping.

%We find that the calculated spectral functions for layers $i=30$ and
%$i=15$ agree well with the surface and bulk spectral functions measured in experiments.\cite{Zabolotnyy2007,Nakayama2007,Okawa2009} 

In Fig.~\ref{fig:simpledos_1}, the intensity of $A_i(\kk,\omega)$
strongly depends on the amount of hybridization between chain and
plane states.  The Fermi surface at the CuO-terminated surface ($i=1$)
is relatively undistorted, indicating weak chain-plane coupling.  By
contrast, the chain Fermi surface in the bulk is more strongly
hybridized to the CuO$_2$ plane states and is correspondingly washed
out.  This is a possible reason that the chain Fermi surface was
imaged in surface-sensitive ARPES
measurements\cite{Zabolotnyy2007,Nakayama2007} but not in
bulk-sensitive measurements.\cite{Okawa2009}

We remark that the amount of hybridization between chains and planes
is strongly $\kk$-dependent, even though the matrix element $t_{\perp
  c}$ is independent of $\kk$.  This is because the hybridization at
each $\kk$ depends on the energy difference between the chain and
plane bands at that value of $\kk$.  Because these bands have
different symmetries, the energy difference (and thus the
hybridization) is a strong function of $\kk$.  It follows that the
amount of hybridization at $\varepsilon_F$ is a strong function of the
relative positions of chain and plane Fermi surfaces, with the
hybridization being largest where the Fermi surfaces cross.  This
explains the difference between the surface ($i=1$) and bulk ($i=13$)
spectral functions shown in Fig.~\ref{fig:simpledos_1}.  A more
extensive discussion of this point can be found in
Ref.~\onlinecite{Atkinson1999}.

The degree to which the chain and plane Fermi surfaces hybridize
determines the size of the induced gap in the chain layers.  The small
gap at the CuO surface (Fig.~\ref{fig:simpledos}) is thus due to the
weak hybridization of the surfaces chains with the underlying CuO$_2$
planes.  In the bulk, the chain DOS shown in Fig.~\ref{fig:simpledos}
has a main gap of about 20 meV, and a small gap of about 4 meV.  The
small gap originates from sections of the chain Fermi surface that are
only weakly coupled to the planes (namely, $k_x\lesssim1$) while the
large gap comes from sections of the Fermi surface that are strongly
coupled to the planes ($k_x\gtrsim1$).  The small gap was discussed
previously as a possible source for the subgap structure measured in
some tunneling experiments.\cite{Ngai2007}

Having discussed general features of the DOS and spectral function, we
now discuss how these are affected by changes in specific model
parameters.  First, we allow for the possibility that the surface
layers of the YBCO thin film are doped by the adsorption of atoms or
molecules onto the surfaces.  Adsorption happens naturally, for
example, when YBCO is exposed to air, and deliberate potassium
adsorption has been used to control the electron concentration in
CuO$_2$ surface states.\cite{Hossain2008} In this work, we are
interested in the possibility that the adsorbate layers partially
screen the electric fields at the YBCO surfaces.  We model this by
assuming that the adsorbed layer has an average charge density of
$+en_s$ ($-en_s$) at the CuO$_2$ (CuO) surface, where $n_s>0$.  In our
self-consistent calculations, the effect of the adsorbate charge is to
reduce the charge transfer between the chain and plane surfaces.

\begin{figure}
\includegraphics[width=\columnwidth]{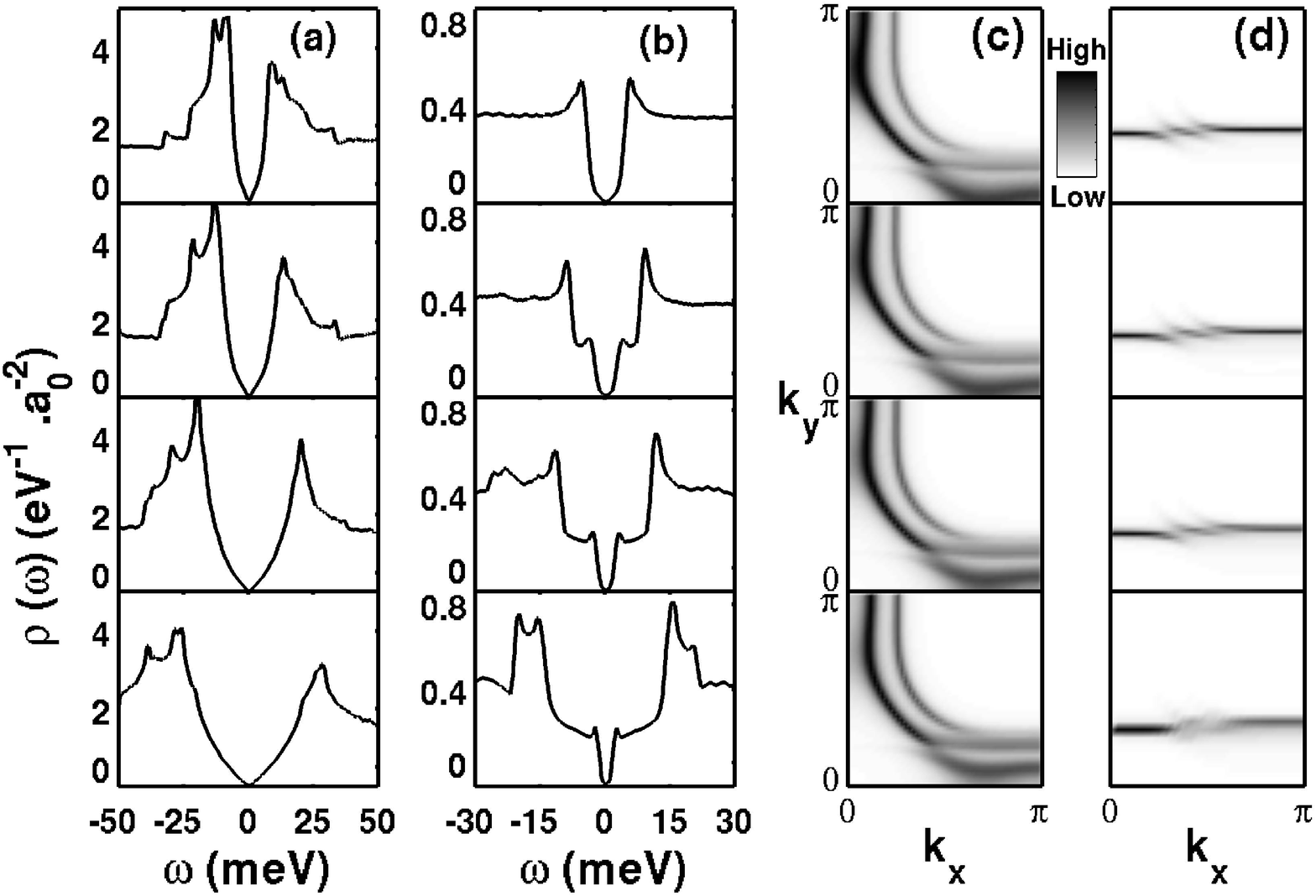}
\caption{Effects of an adsorbed surface layer on the surface states of
  $\mathrm{YBa_2Cu_3O_{6.92}}$.  Columns are for (a) plane and (b)
  chain DOS, (c) plane and (d) chain spectral function.  The adsorbate
  layer has a 2D charge density of $+en_s$ for the CuO$_2$ surface and
  $-en_s$ for the CuO surface.  Rows correspond to $n_s= 0.0$, 0.1,
  0.15, and 0.2 (top to bottom). }
\label{fig:surfacecharge_1}
\end{figure}

Figure \ref{fig:surfacecharge_1} shows the effect of $n_s$ on the
density of states and spectral function.  We see that even a
relatively small adsorbate charge density has a significant effect on
both $\rho_i(\omega)$ and $A_i(\kk,\omega)$ at the surfaces.  In
particular, $\rho_i(\omega)$ and $A_i(\kk,\omega)$ at the surfaces are
increasingly similar to the bulk as $n_s$ increases.  This follows
directly from the fact that the charge densities at the surfaces
approach their bulk values as $n_s$ increases.  In the CuO$_2$ layers,
this results in a larger $\Delta_{id}$ from Eq.~(\ref{eq:simple_gap}),
while in the CuO layers, this results in an increased hybridization
between the surface chains and the adjacent CuO$_2$ plane.

The CuO$_2$ spectrum for $n_s=0.15$ shown in 
Fig.~\ref{fig:surfacecharge_1} is consistent with
existing tunneling experiments on optimally-doped YBCO.
For this case, we obtain coherence peaks at the CuO$_2$ surface at $\pm
20$~meV and a satellite peak at $\approx -30$~meV, in approximate
agreement with
Refs.~\onlinecite{Maggio-Aprile1995,CrenEPL2000,Beyer2009} (when
comparing with experiments, recall that a peak at negative energy in
the DOS corresponds to a peak at positive voltage bias in a tunneling
experiment).  We note that having a charged adsorbate layer is not the
only way to obtain agreement with experiments; a different value for
the dielectric constant, for example, will affect the surface charge
density, and consequently the DOS.  Whether or not our model is
correct in all details, it demonstrates that existing tunneling
measurements on optimally-doped YBCO can be explained within a band
picture.

\begin{figure}
\includegraphics[width=\columnwidth]{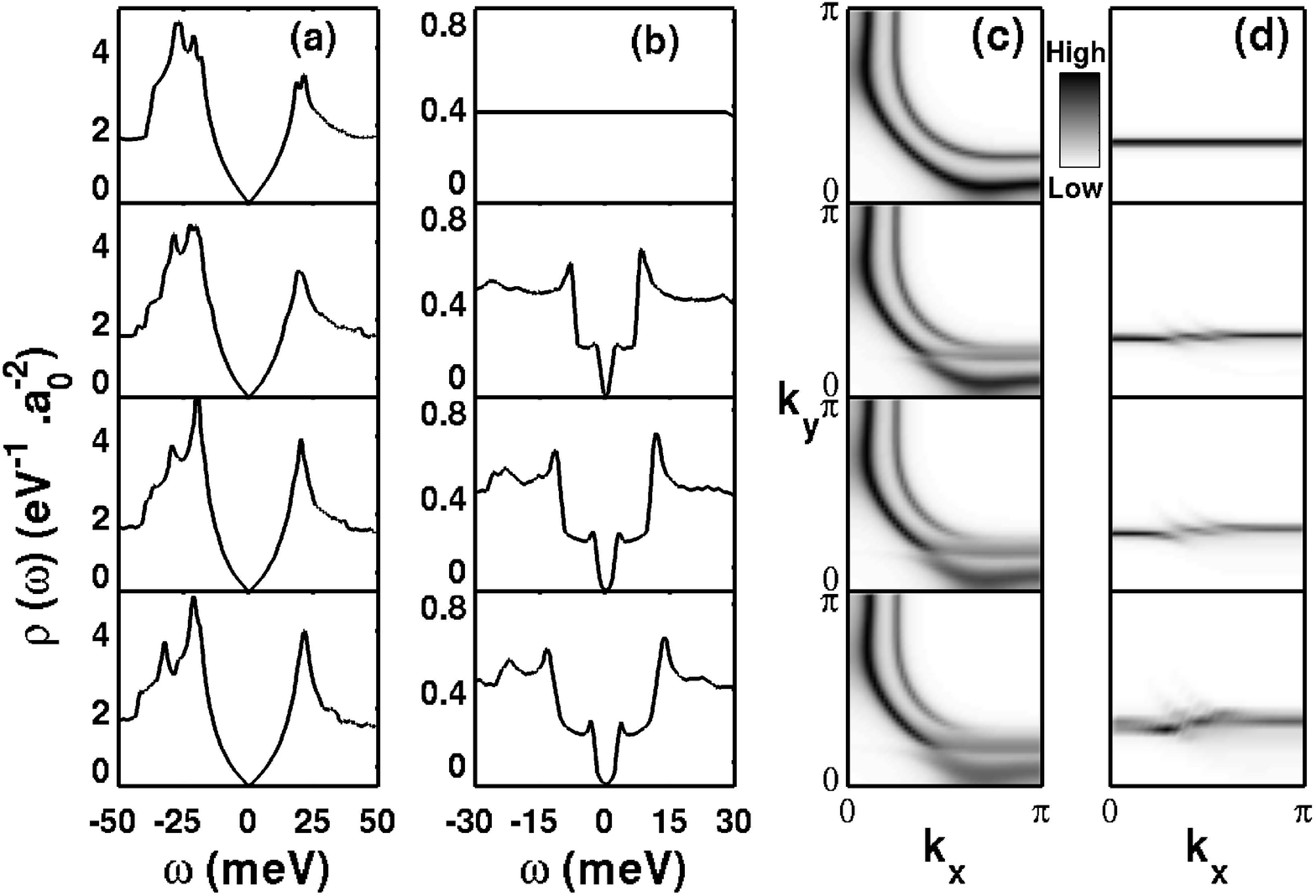}
\caption{Effects of chain-plane coupling on the surface states of
  YB$_2$Cu$_3$O$_{6.92}$.  Columns are for (a) plane and (b) chain DOS
  and (c) plane and (d) chain spectral function.  The adsorbate charge
  density is $n_s = 0.15$.  Rows correspond to $t_{\perp c}= 0.0,\,
  0.8t_{\perp p},\, 1.1t_{\perp p},\, 1.3t_{\perp p}$ (top to
  bottom).}
\label{fig:tpc_2}
\end{figure}

Throughout this work, we have assumed that the plane-chain coupling
parameter is $t_{\perp c}=1.1t_{\perp p}=67$~meV.  This is the hardest
of the model parameters to establish, and was chosen because it gives
reasonable results for $\rho_i(\omega)$ and $A_i(\kk,\omega)$.  Figure
\ref{fig:tpc_2} shows the effect of varying $t_{\perp c}$ on the
density of states and spectral function.  Not surprisingly,
chain-plane coupling has little effect on the DOS at the CuO$_2$
surface, apart from a shift of a weak negative-energy satellite peak
away from the Fermi energy with increasing $t_{\perp c}$.  In
contrast, the chain surface is strongly influenced by coupling to the
CuO$_2$ layer; the CuO chains are metallic when $t_{\perp c}=0$, and
an induced gap appears when $t_{\perp c}$ is nonzero.  The induced gap
grows approximately linearly with $t_{\perp c}$ and has both a subgap
and a main gap, as discussed earlier.  The spectral function also
changes with increasing $t_{\perp c}$, becoming increasingly distorted
near $(\pi,0)$.

%----------------- effect of Ca on the LDOS and ARPES---------------------------------------
\begin{figure}
\includegraphics[width=\columnwidth]{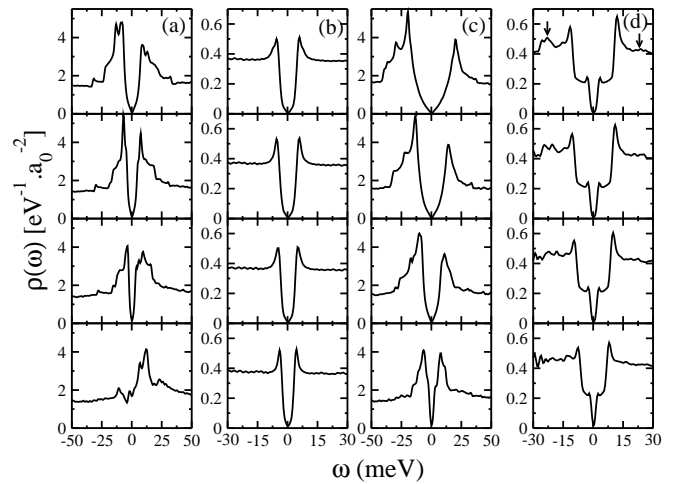}
\caption{Effects of Ca doping on the density of states of \YCBCOX{} at
  the surfaces.  Columns are for (a) plane and (b) chain surfaces with
  $n_s = 0.0$, and (c) plane and (d) chain surfaces with $n_s = 0.15$.
  Rows correspond to $y= 0.0,\, 0.05,\, 0.10,\, 0.2$ (top to
  bottom). Arrows in (d) indicate the locations of satellite features
  discussed in the text.}
\label{fig:change_ca}
\end{figure}

\begin{figure}
\includegraphics[width=\columnwidth]{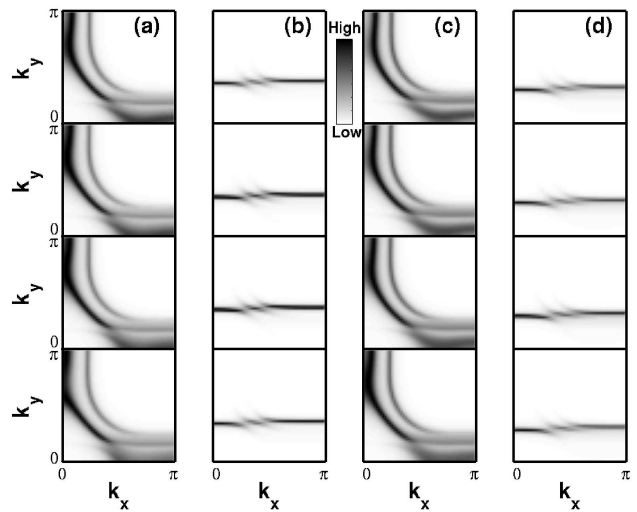}
\caption{Effect of Ca doping on the spectral function at the surfaces of 
\YCBCOX.  Columns are for $n_s=0.0$ for (a) planes and (b) chains, and for
$n_s=0.15$ for (c) planes and (d) chains. Rows are for $y= 0.0,\, 0.05,\, 0.10,\, 0.2$ (top to bottom).}
\label{fig:ARPES}
\end{figure}

Finally, we look more closely at the effects of Ca doping.  In Ngai et
al,\cite{Ngai2007} samples with up to 20\% Ca substitution for Y were
studied, while in Yeh et al\cite{Yeh2001}, 30\% Ca-doped samples were
studied.  In Fig.~\ref{fig:change_ca}, we show a series of
calculations for the surface DOS of a \YCBCOX{} thin film.  Results
are shown for both $n_s=0$ and $n_s=0.15$.  The corresponding spectral
functions are shown in Fig.~\ref{fig:ARPES}.

As discussed above, the DOS exhibits subgap, main gap, and satellite
features.  In Ngai
et al, the three features were attributed to superconductivity on
different regions of the Fermi surface. This is generally consistent
with our findings here although we have found it difficult to
attribute some satellite peaks to specific Fermi surface elements.
Experimentally, the ratio between the satellite, main gap, and subgap
energies was found to be approximately constant as a function of
Ca doping, and this was argued to show that there is a common pairing
mechanism for all three features.  In our model the energies of the
satellite and subgap features in Fig.~\ref{fig:change_ca} also scale
with the main gap.

In some cases, our calculations reproduce the detailed structure of
the experimental spectra.  For $y=0.05$, Ngai et al showed two types
of spectrum.  The first is remarkably similar to that shown for
$y=0.05$ in Fig.~\ref{fig:change_ca}(c), having pronounced coherence
peaks and satellite features resembling shoulders.  Spectra of this
type are also measured in optimally-doped ($y=0$)
samples.\cite{Maggio-Aprile1995,Ngai2007,CrenEPL2000,Beyer2009} The
good agreement between our calculations and the measured spectra
suggests that our model captures the essential physics of the surface
states in optimally-doped YBCO.
 
The second type of measured spectrum qualitatively resembles that
shown for $y=0.0$ in Fig.~\ref{fig:change_ca}(d), having weak
coherence peaks, a subgap feature, and satellite peaks (indicated by
arrows in Fig.~\ref{fig:change_ca}).  However, there is a discrepancy
between our calculations and the experiments; at higher Ca doping
levels, the measured spectra continue to exhibit three sets of peaks
while the satellite peaks in our calculations become less prominent.
Given the sensitivity of the DOS to small changes in the model
parameters, it is plausible that this discrepancy can be corrected by
small changes to the model.  It is also possible that extrinsic
effects not considered here, for example tunneling matrix elements
that emphasize the $(\pi,0)$ and $(0,\pi)$ regions of the Brillouin
zone,\cite{Martin2002,Nieminen2009} could increase the prominence of
the satellite features in tunneling experiments.  However, we also
cannot rule out the possibility that our simple model for Ca
substitution is overly na\"ive.

%It is possible that the second type of spectrum comes from CuO chain terminated surfaces, in which case the subgap and main gap are induced by the proximity effect as described above, and  it remains to understand how the splitting of the main-gap coherence peaks comes about.  Such splitting does occur in our calculations when we assume larger values for $n_s$ (see, for example, Fig.~\ref{fig:surfacecharge_1} for $n_s=0.2$), but is not a robust feature of the model.  
 
%Alternatively, the measured spectra might come from CuO$_2$-terminated surfaces, in which case the coherence-peak splitting can be understood in our model (it occurs frequently in our calculations);   it is the subgap feature which is difficult to explain.  In this case, the subgap might correspond to an impurity resonance associated with scattering from the Ca impurities, or to the onset of a subdominant order parameter which appears upon overdoping.\cite{Yeh2001}  Both of these interpretations have problems, though.  One would expect the width of the subgap peaks to depend on Ca concentration if the peaks were impurity resonances, and one would expect the subgap energy to be independent of the main gap energy if the subgap were due to  subdominant order.\cite{Ngai2007}

Another feature of the DOS that is not explained by our model is the
large residual DOS measured 
experimentally; the DOS is never seen to vanish at $\varepsilon_F$ in
the superconducting state, and is often 50\% of the normal state DOS.
It is not clear whether the residual DOS is intrinsic (for example,
due to pair breaking at the surface) or extrinsic (coming from surface
states in an adsorbate layer).  It is possible that a full description
of the YBCO surface states will require a proper accounting of this
residual DOS.

%------------------------------Conclusions--------------------------------------------------
\section{conclusions}

We have studied the surface and bulk states of \YCBCOX{} with
$0<y<0.2$ within a tight-binding model.  The model parameters for the
CuO$_2$ planes are extracted from photoemission experiments, and
self-consistent calculations are used to relate the surface and bulk
states.  We have calculated the density of states $\rho_i(\omega)$ and
spectral function $A_i(\kk,\varepsilon_F)$ as functions of adsorbed
surface charge density, chain-plane coupling, and Ca doping.  Our main
findings are that
\begin{itemize}
\item our model produces results which are in simultaneous agreement
  with surface and bulk ARPES measurements.  This supports two
  key assumptions of the model, that the surface and bulk bands are
  connected by a simple chemical potential shift and that the Coulomb
  potential can be treated in a planar approximation.
\item the DOS in optimally-doped YBCO can be quantitatively explained
  by our model.  In particular, shoulders measured in the density of
  states that were previously attributed to pairing or to competing
  order are found to be band structure effects.
\item the superconducting DOS is sensitive to small changes in the
  model parameters.  This suggests that, in materials with CuO chains,
  small changes in doping of the surface states can have a qualitative
  effect on the density of states, purely as a result of changes to
  the band structure.  This should be contrasted with BSCCO where it
  is believed that the doping dependence of the DOS is primarily due
  to strong correlations.
\item we can understand some features of Ca-doped YBCO; for example,
  our calculations find densities of states at the chain surface with
  subgap, main gap, and satellite features similar to experiments.
  However, we have not understood the dependence of this spectrum on
  Ca concentration.
\end{itemize}

\section*{Acknowledgments}
We would like to acknowledge helpful coversations with J. Mannhart,
T. Kopp, J. Wei, and J. Ngai. K.P. thanks The Commission on 
Higher Education Grant, Thailand (Grant No.12/2548) 
for financial support. This work was supported by NSERC 
of Canada and by SFB 484 from the DFG.  This work was, in part, made possible by the facilities of the Shared Hierarchical Academic Research Computing Network (SHARCNET:www.sharcnet.ca) and Compute/Calcul Canada.

\bibliographystyle{apsrev}
%\bibliography{ybco}
%copy file .bbl to plate here

\end{document}